\documentclass{aa}  

\usepackage[switch]{lineno}

\usepackage{stfloats}
\usepackage{placeins}
\usepackage{afterpage}

\usepackage{graphicx}

\usepackage{txfonts}
\usepackage{enumitem}
\usepackage{multirow}
\usepackage{booktabs}

\usepackage{natbib,twoopt}
\usepackage[breaklinks=true, colorlinks=true, citecolor=blue, urlcolor=blue]{hyperref} 

\bibpunct{(}{)}{;}{a}{}{,}       
\makeatletter
  \newcommandtwoopt{\citeads}[3][][]{\href{http://adsabs.harvard.edu/abs/#3}
    {\def\hyper@linkstart##1##2{}
     \let\hyper@linkend\@empty\citealp[#1][#2]{#3}}}
  \newcommandtwoopt{\citepads}[3][][]{\href{http://adsabs.harvard.edu/abs/#3}
    {\def\hyper@linkstart##1##2{}
     \let\hyper@linkend\@empty\citep[#1][#2]{#3}}}
  \newcommandtwoopt{\citetads}[3][][]{\href{http://adsabs.harvard.edu/abs/#3}
    {\def\hyper@linkstart##1##2{}
     \let\hyper@linkend\@empty\citet[#1][#2]{#3}}}
  \newcommandtwoopt{\citeyearads}[3][][]
    {\href{http://adsabs.harvard.edu/abs/#3}
    {\def\hyper@linkstart##1##2{}
     \let\hyper@linkend\@empty\citeyear[#1][#2]{#3}}}
\makeatother

\titlerunning{Leveraging pre-trained vision Transformers for multi-band photometric light curve classification}
\begin{document} 
   \title{Leveraging pre-trained vision Transformers for multi-band photometric light curve classification}

   \author{
          D. Moreno-Cartagena
          \inst{1,2}\textsuperscript{\tiny\thanks{e-mail: \href{mailto:dmoreno2016@inf.udec.cl}{\texttt{dmoreno2016@inf.udec.cl}}}}
          \and
          P. Protopapas\inst{3}
          \and
          G. Cabrera-Vives\inst{1,2,4,5}
          \and
          M. Cádiz-Leyton\inst{1,7} 
          \and
          I. Becker\inst{3}
          \and
          C. Donoso-Oliva\inst{1,6}
          }

   \institute{
    Department of Computer Science, Universidad de Concepción, Edmundo Larenas 219, Concepción, Chile
    \and
    Center for Data and Artificial Intelligence, Universidad de Concepción, Edmundo Larenas 310, Concepción, Chile
    \and
    John A. Paulson School of Engineering and Applied Science, Harvard University, Cambridge, MA, 02138
    \and
    Heidelberg Institute for Theoretical Studies, Heidelberg, Baden-Württemberg, Germany
    \and
    Millennium Institute of Astrophysics (MAS), Nuncio Monseñor Sotero Sanz 100, Of. 104, Providencia, Santiago, Chile
    \and
    Millennium Nucleus on Young Exoplanets and their Moons (YEMS), Chile
    \and
    Edinburgh Futures Institute, University of Edinburgh, 1 Lauriston Pl, Edinburgh EH3 9EF, UK
    }

   \date{Received September 15, 1996; accepted March 16, 1997}

  \abstract  
   {   
    The advent of large-scale sky surveys, such as the Vera C. Rubin Observatory Legacy Survey of Space and Time (LSST), will generate vast volumes of photometric data, necessitating automatic classification of light curves to identify variable stars and transient events. However, challenges such as irregular sampling, multi-band observations, and diverse flux distributions across bands demand advanced models for accurate classification. 
   }
   {
    This study investigates the potential of a pre-trained vision Transformer (VT) model, specifically the Swin Transformer V2 (SwinV2), to classify photometric light curves without the need for feature extraction or multi-band preprocessing. The goal is to assess whether this image-based approach can accurately differentiate astronomical phenomena and if it can serve as a viable option for working with multi-band photometric light curves.
   }
   {
    We transformed each multi-band light curve into an image. These images served as input to the SwinV2 model, which was pre-trained on ImageNet-21K. The datasets employed include the public Catalog of Variable Stars from the Massive Compact Halo Object (MACHO) survey, using both one and two bands, and the first round of the recent Extended LSST Astronomical Time-Series Classification Challenge (ELAsTiCC), which includes six bands. The model’s performance was evaluated based on six classes for the MACHO dataset and 20 distinct classes of variable stars and transient events for the ELAsTiCC dataset.
   }
    {
    The fine-tuned SwinV2 model achieved a better performance than models specifically designed for light curves, such as Astromer and the Astronomical Transformer for time series And Tabular data (ATAT). When trained on the ``full dataset'' of MACHO, it attained a macro F1-score of 80.2\% and outperformed Astromer in single-band experiments. Incorporating a second band further improved performance, increasing the F1-score to $84.1\%$. In the ELAsTiCC dataset, SwinV2 achieved a macro F1-score of $65.5\%$, slightly surpassing ATAT by $1.3\%$.
    }
    {
    SwinV2, a pre-trained VT model, effectively classifies photometric light curves. It outperforms traditional models and offers a promising approach for large-scale surveys. This highlights the potential of using visual representations of light curves, with future prospects including the integration of tabular data, textual information, and multi-modal learning to enhance analysis and classification in time-domain astronomy.
    }

   \maketitle

\section{Introduction}
The Vera C. Rubin Observatory’s Legacy Survey of Space and Time \citep[LSST;][]{ivezic2019lsst} is set to revolutionize our understanding of the Universe by capturing an unprecedented volume of data, with an average of 10 million alerts and $\sim 20\,\text{TB}$ of raw data every night \citep{LSST:2022kad}. This vast dataset covers a range of celestial phenomena, including transient events and stellar variability, and will provide astronomers with a unique opportunity to study the physical properties and evolution of a diverse array of astronomical objects. However, the scale of these observations necessitates the development of efficient and accurate classification tools, as spectroscopic follow-up will be limited. Automatic classification methods, particularly those based on photometric data, will be essential for distinguishing between different types of variable stars and transient phenomena, ensuring that the scientific potential of the LSST data is fully realized \citep{fraga2024transient}.

Machine learning and deep learning techniques have played a central role in advancing the classification of light curves \citep{pichara2012improved, kim2011quasi, villar2019supernova, becker2020scalable, cabrera2024atat}. Traditional machine learning approaches have been highly effective in classifying variable stars and transient events by relying on hand-engineered features extracted from light curves. For example, \citet{richards2011machine} and \citet{kim2014epoch} employed statistical and time-domain features to enhance classification accuracy. \citet{karpenka2013simple} combined parametric functional fitting of supernova light curves with a machine learning algorithm, yielding accurate and robust classifications. \citet{lochner2016photometric} developed a multi-stage pipeline that evaluates feature extraction techniques and machine learning models. \citet{boone2019avocado} improved classification by training a boosted decision tree on features extracted from the augmented light curves, achieving state-of-the-art performance in the LSST Astronomical Time-Series Classification Challenge \citep[PLAsTiCC;][]{kessler2019models}. However, this feature engineering process is resource-intensive and requires domain knowledge to be applied \citep{donalek2013feature, graham2014novel, nun2015fats, sanchez2021alert}. Deep learning addresses these limitations by automatically learning relevant representations directly from light curves, removing the need for manual feature extraction. This advantage has led to the widespread adoption of recurrent neural networks \citep[RNNs;][]{rumelhart1986learning}, long short-term memory networks \citep[LSTMs;][]{hochreiter1997long}, and gated recurrent units \citep[GRUs;][]{cho-etal-2014-learning} \citep{charnock2017deep, naul2018recurrent, muthukrishna2019rapid, carrasco2019deep, moller2020supernnova, becker2020scalable, jamal2020neural, gomez2020classifying, donoso2021effect} as well as transformer-based models, which have shown great promise in handling astronomical time series data \citep{morvan2022don, pan2022astroconformer, pimentel2022deep, donoso2023astromer, moreno2023positional, cadiz2024workshop, cadiz2024Journal, allam2024paying, cabrera2024atat}, making them increasingly popular for astronomical classification tasks. 

Nevertheless, even with deep learning models in the time domain, careful preprocessing is required when dealing with multi-band observations and irregular time sampling, as each band is often observed at different non-uniform time intervals. Various techniques have been developed to address these challenges, such as uniform sampling imputation between previous and subsequent observations \citep{charnock2017deep}, linear interpolation on a regular time grid across all bands \citep{muthukrishna2019rapid}, Gaussian process (GP) interpolation \citep{boone2019avocado, villar2020superraenn, jamal2020neural}, masking mechanisms to indicate the availability of each band during observations \citep{moller2020supernnova, pimentel2022deep, cabrera2024atat}, and ensembles of RNNs that integrate information from different bands into a single representation \citep{Becker2025}. While these methods have been evaluated on various datasets, their generalization across different surveys and observational conditions remains an open question \citep{moreno2023positional}. Additionally, many of these approaches require training from scratch and hyperparameter tuning to search for the optimal architecture for a given task.

Foundational models, which are trained only once on massive datasets in a self-supervised manner (pre-training) to learn general domain representations and later adapted for specific tasks (fine-tuning), have demonstrated exceptional generalization across diverse downstream applications in natural language processing and computer vision, such as text generation \citep{brown2020language}, machine translation \citep{vaswani2017attention}, image classification \citep{dosovitskiy2021an}, and object detection \citep{carion2020end}, among others. Notable examples include GPT \citep{radford2018improving} and BERT \citep{devlin2019bert} for text as well as Vision Transformer \citep[ViT;][]{dosovitskiy2021an} for images, with each model specializing in a single modality. In the context of astronomical time series data, Astromer \citep{donoso2023astromer} represents an early effort to pre-train a transformer on large-scale real-world light curve data, positioning it as a potential foundational model for time-domain astronomy. However, such pre-training requires significant computational resources, optimized architectural design, and carefully curated datasets, often taking several days to converge. 

Given these constraints, we propose leveraging the knowledge embedded in pre-trained vision Transformer (VT) models for the analysis and classification of multi-band photometric light curves. These models have demonstrated the ability to capture fine details in images across various tasks \citep{han2023survey, khan2022survey}. By using the informative representations learned from extensive pre-training on large-scale datasets conducted on powerful GPU clusters, we can establish a starting point for fine-tuning the model on light curve images, bypassing the need to use significant computational resources for pre-training and searching for an optimal architecture. Moreover, these models can directly process raw multi-band light curves as images, offering a straightforward yet highly effective approach to handling the data while avoiding complex feature engineering or multi-band preprocessing.

A few studies have considered light curves in the time domain for visual models. \citet{mahabal2017deep} transformed light curves from the Catalina Real-time Transient Survey \citep[CRTS;][]{drake2012catalina} into 2D images by creating magnitude differences and time differences between pairs of points as input for a convolutional neural network \citep[CNN;][]{lecun1998gradient}. \citet{pasquet2018deep} and \citet{pasquet2019pelican} used CNNs for the classification of astronomical events, representing light curves as images, where the vertical axis corresponds to the number of photometric bands and the horizontal axis represents the number of days in the first study and the number of observations in the second. \citet{qu2021scone} and \citet{qu2022photometric} transformed multi-band light curves into 2D heatmaps using GPs to represent flux and its uncertainties over time and wavelength. These heatmaps were stacked into $32 {\times} 180 {\times} 2$ tensors and fed into a CNN for supernova classification. \citet{szklenar2020image, szklenar2022variable} explored the classification of phase-folded light curves using the Optical Gravitational Lensing Experiment (OGLE)-III and -IV databases \citep{Udalski2008, Udalski2015}, representing them as $128 {\times} 128$ monochromatic pixel images with brightness measurements plotted as white dots on a black background. Similarly, \citet{monsalves2024application} transformed light curves into $32 {\times} 32$ pixel 2D histograms, where phase and magnitude bins were used as inputs to a CNN to classify different types of variable stars. However, all of these studies depended on a prior level of preprocessing and did not use pre-trained VT models.

In this work, we present the use of a pre-trained VT model called Swin Transformer V2 \citep[SwinV2;][]{liu2022swin}, which we evaluate on two datasets: the public Catalog of Variable Stars \citep{AlcockData}, derived from the Massive Compact Halo Object \citep[MACHO;][]{MACHOData} survey, and the first round of the recent Extended LSST Astronomical Time-Series Classification Challenge \citep[ELAsTiCC\footnote{\href{https://portal.nersc.gov/cfs/lsst/DESC_TD_PUBLIC/ELASTICC/}{The DESC ELAsTiCC Challenge.}};][]{Narayan2023Extended}, a multi-band dataset of photometric light curves for variable stars and supernovae designed to simulate the observations of the future LSST. Our work builds on the methodology proposed by \citet{li2024time}, which focuses on benchmark datasets for time series. We applied this method to light curves, performed hyperparameter tuning, and evaluated its performance using both single- and multi-band configurations for the MACHO dataset and six bands for the ELAsTiCC dataset. Additionally, we assessed its performance in scenarios with a limited number of astronomical objects per class and imbalanced datasets. To our knowledge, this is the first study to apply a pre-trained VT to raw multi-band light curves in the time domain. 

This paper is organized as follows: In Section~\ref{sec:methods}, we describe the methodology for converting multi-band light curves into images and present the architecture of the SwinV2 vision Transformer model. Section~\ref{sec:experimental_setup} provides details of the datasets and experimental setup. In Section~\ref{sec:results}, we present the results of our classification experiments on the MACHO and ELAsTiCC datasets, including comparisons with established methods. Finally, in Section~\ref{sec:conclusions} we summarize the main conclusions of this work and suggest directions for future research.

\section{Methods}\label{sec:methods}

\subsection{Multi-band light curves to images}
\begin{figure}[t]
  \centering
  \includegraphics[width=0.66\hsize]{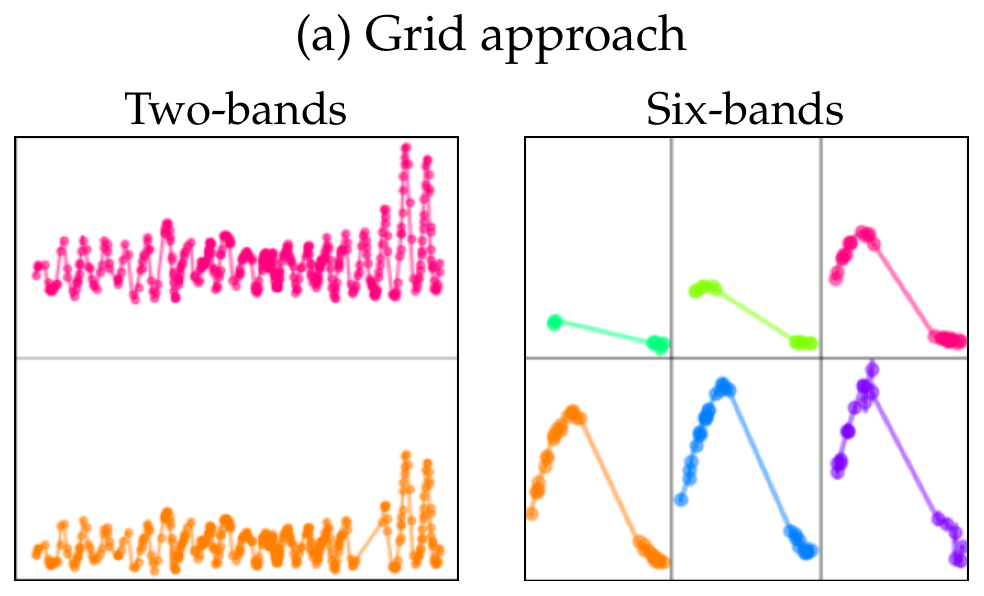} \\
  \vspace{2.5pt}
  \includegraphics[width=\hsize]{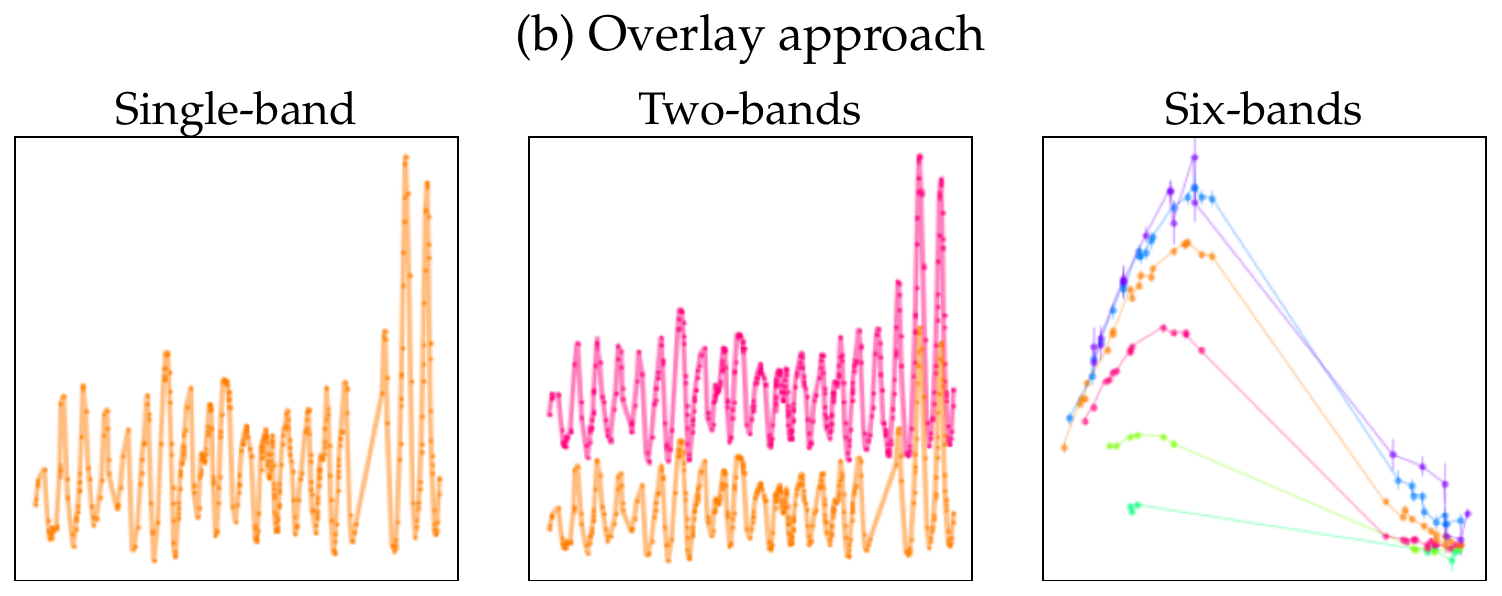}
  \caption{Comparison of visualization strategies. Panel (a) illustrates the ``Grid'' approach, while panel (b) depicts the ``Overlay'' approach. Each image was generated using the best hyperparameters identified for its respective method. The single-band and two-band example is a Long-Period Variable (LPV) (ID: 24.3466.13) from the MACHO dataset, whereas the six-band example is a Pair-Instability Supernova (PISN) (ID: 41820707) from the ELAsTiCC dataset.}
  \label{fig:input_model}
\end{figure}

Transforming time series into images is crucial for applying VT models, which have proven effective in analyzing complex visual data \citep{dosovitskiy2021an, liu2021swin, liu2023survey}. In this study, we evaluate two approaches for converting multi-band light curves into images: ``Grid'' and ``Overlay''. Figure~\ref{fig:input_model} (a) shows the input for the ``Grid'' approach, while (b) illustrates the input for the ``Overlay'' approach.

In the ``Grid'' approach, light curves from different bands are arranged in a grid, with each band displayed in a separate panel, following a similar methodology to \citet{li2024time}. We use a square layout that adjusts to the number of bands present in each light curve, leaving empty panels when there are no observations for a particular band. This arrangement facilitates a clear comparison between bands by keeping them visually distinct. In the ``Overlay'' approach, all bands of a light curve are plotted together within the same space. This approach allows us to examine how interactions among observations from different bands influence the visual representation and model performance.

To ensure that all light curve images are on the same scale and maintain consistent x- and y-limits across different images, we applied min-max normalization to each entire light curve (considering all bands). This approach preserves the relative differences among bands within each light curve while scaling time and flux values across all light curves to the range [$0$, $1$].  As a consequence, the global temporal and brightness context across light curves is removed, which encourages the model to focus on light curve shape to discriminate between astronomical objects. Global normalization was avoided, as it suppressed low-amplitude variability and negatively affected classification performance. Axis labels and ticks are also removed before saving the plots. Additionally, the flux errors are scaled using the same Min-Max factor applied to the flux, maintaining the relative uncertainty information. This method ensures that all images are visually comparable, with uniform axis limits and scale throughout the dataset. Visual examples of these representation approaches applied to randomly selected light curves from the test set are provided in Appendix~\ref{appendix:visual_samples}, in Fig.~\ref{fig:random_samples_macho} and Fig.~\ref{fig:random_samples_elasticc}, offering a broader view of the differences between the One-band, ``Overlay'', and ``Grid'' methods across both the MACHO and ELAsTiCC datasets.

Since pre-trained VT models are trained on RGB images, the input must conform to this expected format, requiring a three-channel representation. In the ``Grid'' approach, the specific color assigned to each band is inconsequential, as each band occupies a distinct spatial position. However, in the ``Overlay'' approach, where all bands are combined into the same visual space, color selection becomes crucial to avoid introducing visual biases that could affect the  model’s performance. To mitigate this issue and ensure that all bands contribute equally to the final representation, we assign colors that approximate distinct RGB components, maintaining a perceptually balanced distribution across the spectrum. For datasets with six bands, we allocate two bands to the red spectrum with the RGB values (255, 0, 127) and (255, 127, 0), two bands to the green spectrum with (0, 255, 127) and (127, 255, 0), and the remaining two to the blue spectrum with (0, 127, 255) and (127, 0, 255). This selection ensures an even distribution of color intensity, preventing any single band from dominating the representation. For datasets containing only two bands, each is mapped to a primary RGB component to preserve visual balance and maintain interpretability. 

Additionally, in both approaches, we consider visual properties as hyperparameters to be adjusted during the conversion of light curves to images. These include marker size (markersize), line width (linewidth), and the inclusion or exclusion of flux uncertainties in the images (yerr). These plot properties influence not only the visualization but also the model's ability to extract relevant features effectively. The visual hyperparameters were optimized to maximize classification performance, as detailed in the implementation subsection.

\subsection{Vision Transformer models}
Vision Transformer models represent a significant evolution in deep learning architectures for image processing. Unlike conventional approaches based on convolutions, such as CNNs, VTs employ self-attention mechanisms to capture long-range relationships between different regions of an image \citep{dosovitskiy2021an}. This methodology enables VTs to capture intricate structures within images, achieving robust performance in various tasks, such as classification \citep{liu2021swin}, object detection \citep{carion2020end}, semantic and instance segmentation \citep{wang2021max, cheng2021per}, object tracking \citep{chen2021transformer}, image generation \citep{jiang2021transgan}, and image enhancement \citep{chen2021pre}, among others. Furthermore, VT models that are pre-trained can use the prior knowledge gained from extensive datasets, making them adaptable to particular tasks via a fine-tuning process.

\begin{figure}[b]
  \resizebox{\hsize}{!}{\includegraphics{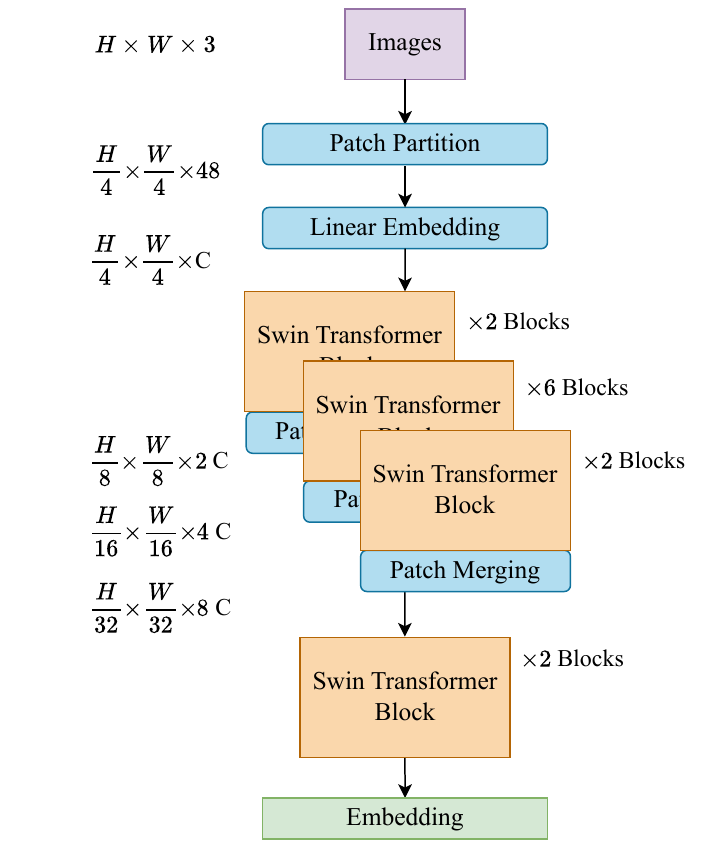}}
  \caption{SwinV2 architecture. The rounded rectangle in light blue highlights the components where the model changes the dimensions of the information, while the regular rectangles indicate where they remain fixed.}
  \label{fig:swin_arch}
\end{figure}

SwinV2 is a hierarchical VT architecture designed for efficient and scalable image processing \citep{liu2022swin}. Unlike ViT \citep{dosovitskiy2021an}, which partitions an image into non-overlapping segments called patches, applying global self-attention to all of them simultaneously, the first Swin Transformer paper \citep{liu2021swin} introduces two key innovations that allow it to manage both local and global self-attention: shifted windows and hierarchical feature representation. These features enable the model to handle high-resolution images with greater computational efficiency. Furthermore, SwinV2 incorporates scaled cosine attention and logarithmic continuous position bias, improving both training stability and performance in tasks such as image classification, object detection, semantic segmentation, and video action classification. We selected this model as the foundation of our approach because its hierarchical structure is particularly effective at capturing detailed information at the pixel level, which is essential for interpreting points and lines within the light curve image. Figure~\ref{fig:swin_arch} illustrates the model's architecture and hierarchical processing. The key stages of this architecture are detailed below.

\begin{figure}[b]
  \resizebox{\hsize}{!}{\includegraphics{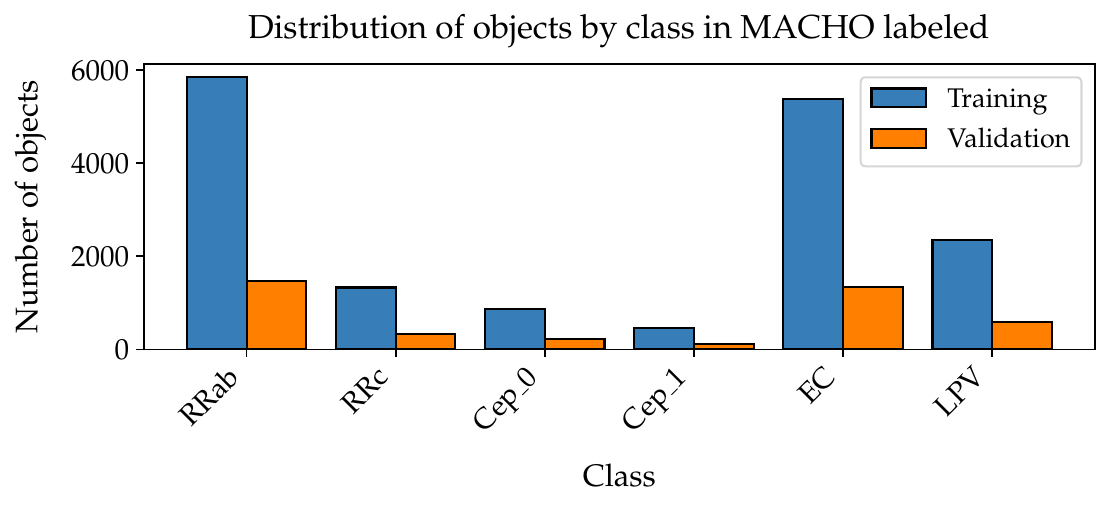}}
  \caption{Distribution of periodic variable star classes in the training and validation sets for the MACHO dataset.}
  \label{fig:distribution_macho}
\end{figure}

\begin{figure*}[b]
    \centering
    \includegraphics[width=17cm]{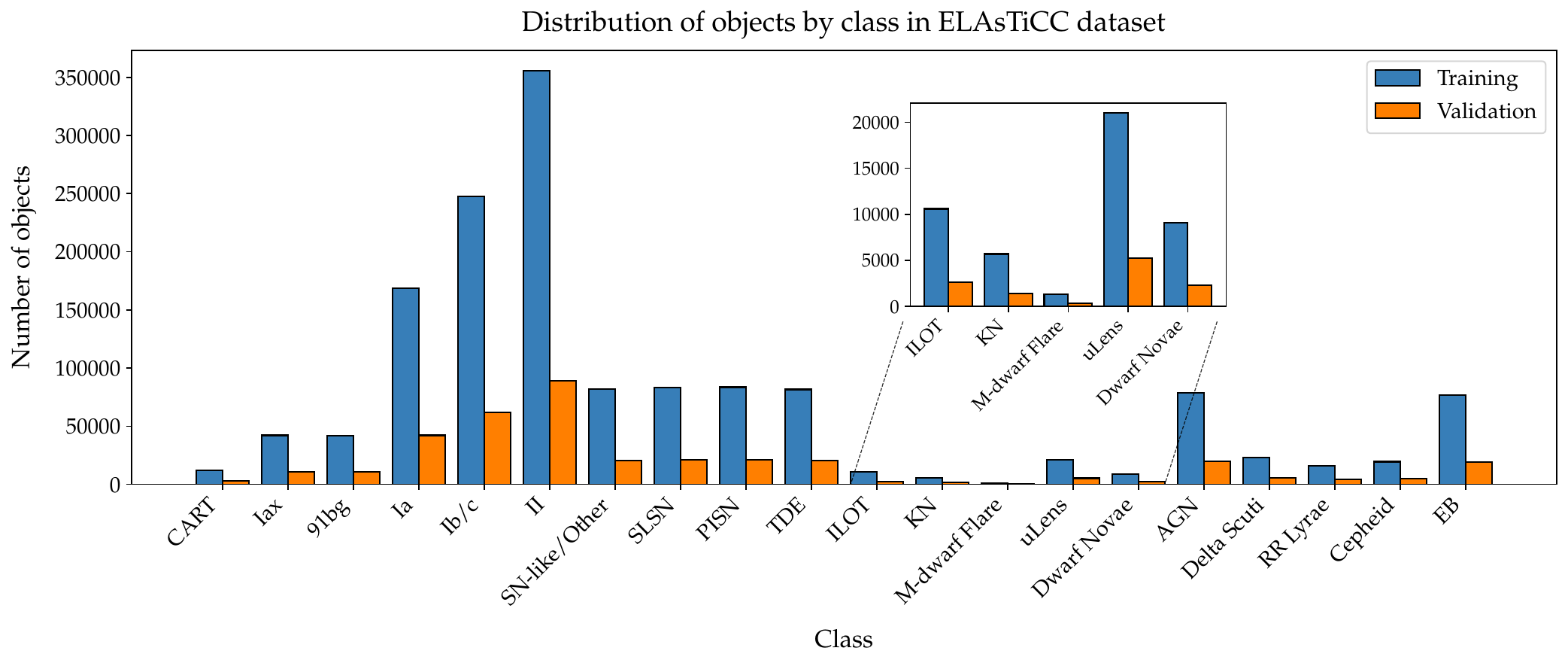}
    \caption{Distribution of transients, stochastic variables, and periodic variable star classes in the training and validation sets for the first round of the ELAsTiCC dataset.}
    \label{fig:distribution_elasticc}
\end{figure*}

The first step in the SwinV2 architecture converts the input image into a sequence of tokens through patch splitting and tokenization. This process begins by partitioning the input image into non-overlapping patches, similar to ViT. However, unlike ViT, Swin progressively reduces the patch size as the image propagates through the Swin Transformer block. Given an input image of \( 256 {\times} 256 \) pixels with 3 color channels (e.g., RGB) and a patch size of \( 4 {\times} 4 \), the patch partitioning process yields \( 64 {\times} 64 \) patches. This is equivalent to applying a convolutional kernel of size \( 4 {\times} 4 \) with a stride of \( 4 {\times} 4 \). Each patch contains $4 {\times} 4 {\times} 3 {=} 48$ values that represent the RGB pixel intensities within that region. These values are then flattened and projected into a higher-dimensional space via a linear transformation, producing a token map of dimensions \( 64 {\times} 64 {\times} C \), where $C$ is the embedding dimension. In this context, a token refers to the vector representation of a patch after the linear transformation. This process is functionally analogous to applying a two-dimensional convolution layer (Conv2d). The resulting token map is then processed through the Swin Transformer block for hierarchical processing.

The window multi-head self-attention (W-MSA) step partitions the tokens into non-overlapping windows of size \(M \times M\). Unlike traditional multi-head self-attention (MSA), which computes attention globally with quadratic complexity, W-MSA restricts attention to local windows. This reduces the computational complexity from $\mathcal{O}(N^2)$ in global attention to $\mathcal{O}(M^2N)$ in W-MSA, where $N$ denotes the total number of tokens in the image. Since $M$ is fixed, W-MSA maintains linear complexity with respect to $N$, significantly reducing computational cost while ensuring that each window effectively captures local dependencies, making the approach scalable for high-resolution images.

To address the limited cross-window interactions in W-MSA, the Shifted Window Multi-head Self-Attention (SW-MSA) mechanism shifts the windows from the W-MSA step, allowing tokens at the edges of one window to be included in the centers of overlapping windows. This process enables information exchange between neighboring windows. By alternating W-MSA and SW-MSA within each Swin Transformer block, the model captures global context while preserving the computational advantages of local attention. For a more detailed explanation of the W-MSA and SW-MSA mechanisms, please refer to the first Swin Transformer paper \citep{liu2021swin}.

Most VT models, including the original Swin Transformer, apply layer normalization before the attention mechanism. However, as model depth increases, deeper layers develop larger activation magnitudes due to residual connections, leading to training instability. To address this, SwinV2 adopts residual post-normalization, where normalization is applied after the attention mechanism and the output of the Swin Transformer block, preventing the accumulation of large activations. Additionally, traditional dot-product attention in large models can cause certain heads to focus excessively on a few pixel pairs, particularly in the residual post-normalization setting. To mitigate this, SwinV2 introduces scaled cosine attention, which replaces dot-product similarity with cosine similarity, normalizing attention scores and improving training stability.

SwinV2 also replaces the relative position bias with a log-spaced continuous position bias. This approach is particularly relevant in hierarchical models such as SwinV2, where the patch merging operation, detailed in the following paragraph, changes the token resolution at each stage of the network. The log-spaced continuous position bias provides a flexible encoding of spatial relationships that adapts naturally as the resolution changes, allowing the model to maintain meaningful attention patterns at multiple scales. In practice, this helps the model assign more attention to neighboring regions (where fine details or local patterns often reside) while remaining robust to global variations in the visual structure of different light curves.

Following these operations, the model applies patch merging to create a hierarchical representation. This reduces the spatial resolution of the token map while increasing its embedding dimension, inspired by how CNNs progressively downsample feature maps while enriching feature abstraction. During patch merging, each group of \(2 \times 2\) neighboring tokens is concatenated to form a single new token, halving the height and width of the token map and multiplying the embedding dimension by four. A linear layer then projects these concatenated features into a lower-dimensional space, typically doubling the previous embedding size (e.g., from \(64 \times 64 \times 96\) to \(32 \times 32 \times 192\)). By repeating this process across network stages, the model builds a multi-scale or hierarchical representation: early layers retain fine-grained spatial details, while deeper layers encode more abstract and global patterns. This hierarchical structure is designed to enable the model to capture features at different temporal scales, which, in the context of light curves, could correspond to both short-term variations and longer-term trends.
   
After the hierarchical stages, the final token representations undergo adaptive average pooling, which computes the mean across all tokens of size $\frac{H}{32} \times \frac{W}{32}$, preserving their aggregated embedding information ($8C$). This process allows each image to be represented as a vector of dimension $8C$, which is then passed through a fully connected linear layer for classification or other downstream tasks.

\section{Experimental setup}\label{sec:experimental_setup}

\subsection{Data description}

\subsubsection{MACHO labeled}
The public Catalog of Variable Stars\footnote{\href{https://vizier.cds.unistra.fr/viz-bin/VizieR?-source=II/247}{MACHO Labeled Data of the Variable Stars.}} \citep{AlcockData} contains labeled light curves from the MACHO survey, which observed variable stars in the Large Magellanic Cloud (LMC) and the Galactic bulge over multiple years, capturing their photometric variability in the B and R bands. These light curves are represented in magnitude space. The catalog labels were assigned through automated classification procedures based on statistical properties of the light curve morphology, such as period, amplitude, and Fourier decomposition coefficients. While visual verification was commonly employed in some of the MACHO team’s publications \citep{alcock1996macho, alcock1997macho_b, alcock1997macho_a, alcock1999macho, alcock2002macho}, there is no evidence in the literature confirming that a systematic visual inspection was performed for the full catalog used in this work. Therefore, we assume that the classification relied solely on the extracted features. It is important to note that we do not use any auxiliary features or pre-computed variables at any stage of the preprocessing or training pipeline, which could otherwise lead to a trivial classification task. This dataset also has a median cadence of $0.75$ days with a standard deviation of $5.03$ days, considering both bands. The labeled portion we use corresponds to the same one used in \citet{donoso2023astromer}, which was categorized into six distinct classes, comprising $20\,894$ variable stars. We refer to this as the ``full dataset''. 

Following the approach in \citet{donoso2023astromer}, we used scenarios with a few samples per class (spc) to train the model and assess its ability to learn from limited labeled data while keeping the test set unchanged. From the ``full dataset'', we extracted 20 spc and 500 spc as our new training datasets. As in the referenced study, the IDs for each fold were selected randomly from the ``full dataset''. For example, the IDs for 20 and 500 spc in the first fold differ from those in the second fold. Our training, validation, and test sets in each fold use the same IDs that were used in the referenced study. The test set contains $100$ light curves per class and remains fixed across all scenarios (20 spc, 500 spc, and ``full dataset''). The training dataset is divided into three stratified folds, maintaining an 80/20 split for training and validation. This results in $96$ and $24$ light curves for the 20 spc scenario, $2\,400$ and $600$ for the 500 spc scenario, and $16\,232$ and $4\,062$ for the ``full dataset''. Figure~\ref{fig:distribution_macho} illustrates the class distribution in the training and validation sets for the ``full dataset''. 

It is worth mentioning that the referenced paper used only the R band. However, in our work, we tested both cases: one using only the R band to provide a direct comparison with Astromer \citep{donoso2023astromer}, a transformer model specifically designed for single-band light curve data, and another incorporating both bands to demonstrate the effectiveness of the multi-band approach.

\subsubsection{ELAsTiCC}
The second dataset used in this study comes from the ELAsTiCC\footnote{\href{https://portal.nersc.gov/cfs/lsst/DESC_TD_PUBLIC/ELASTICC/TRAINING_SAMPLES/}{ELAsTiCC Challenge Data Portal.}} challenge \citep{Narayan2023Extended}, which is designed to simulate the observations of the future LSST at the Vera C. Rubin Observatory. It contains $1\,845\,146$ simulated light curves, and each is represented across six bands (u, g, r, i, z, and Y) covering the visible and near-infrared spectrum. The data includes multiple observations of each object spread over several simulated years. ELAsTiCC spans a diverse array of transient and variable stellar events, categorized into $32$ distinct classes, with a median cadence of $0.93$ days and a standard deviation of $16.18$ days, considering all six bands. Specifically, we used the first ELAsTiCC campaign, which was used in \citet{cabrera2024atat}, and adopted the same class distribution as their study, encompassing $20$ different categories. Additionally, similar to the approach outlined in the referenced study, we used the light curves in flux space and applied the PHOTFLAG\footnote{\href{https://portal.nersc.gov/cfs/lsst/DESC_TD_PUBLIC/ELASTICC/TRAINING_SAMPLES/A_FORMAT.TXT}{Description of the data variables from the first round of ELAsTiCC.}} key to filter out saturated data, which also helps exclude spurious detections outside the main event, particularly for transient objects. For clarity, we note that no derived features from the light curves or metadata were used at any stage of the training process. For each light curve, we also considered both alerts and forced photometry points, beginning $30$ days prior to the first alert up to the final detection. Non-detections following the last detection were omitted from the dataset. 

Our training, validation, and test sets were the same as those used in \citet{cabrera2024atat}. The test set contains $1\,000$ light curves from each class, ensuring a balanced representation. The remaining data is divided into five stratified folds, maintaining an 80/20 ratio for training and validation. This results in $1\,460\,117$ and $365\,029$ light curves in the training and validation sets, respectively. Figure~\ref{fig:distribution_elasticc} shows the distribution of the classes in the training and validation sets. 

\subsection{Implementation details}
The conversion of light curves to images was performed using the Matplotlib Python package, and the images were directly transformed into tensors with RGB channels. We generated images with dimensions of $256 {\times} 256$, as this is the size for which the VT model was pre-trained. In both the ``Grid'' and ``Overlay'' approaches, the images were adjusted to minimize extra space, and in the ``Grid'' approach, borders were added to separate the bands (see Fig.~\ref{fig:input_model} for an illustration of these representations).

In our experiments, we used the SwinV2 model pre-trained on the ImageNet-21K dataset, available through the Hugging Face Python library\footnote{\href{https://huggingface.co/microsoft/swinv2-tiny-patch4-window16-256}{Pre-trained Hugging Face model used.}}. Image processing was performed using the AutoImageProcessor class associated with the pre-trained model, which included resizing to $256 {\times} 256$ pixels, rescaling pixel values to the range $[0,1]$, and normalizing them using the mean and standard deviation from ImageNet-21K. This process ensured the proper adaptation of the input data to the format required by the SwinV2. The model operates on input images of size $256 {\times} 256 {\times} 3$ pixels, divides them into patches of $4 {\times} 4$ pixels, uses an embedding dimension of $C=96$, and performs self-attention within windows of $16 {\times} 16$ patches. The training was conducted on an NVIDIA A100 GPU. 

We experimented with varying key hyperparameters during the conversion of light curves into images to identify the best visual representation that enhances classification performance, as measured by the F1-score, which balances how accurately the model identifies light curves of a given class (precision) and its ability to recognize all light curves that truly belong to that class (recall). Table~\ref{table:hyperparameters} summarizes the hyperparameters used. Since the MACHO labeled dataset contains a limited number of light curves and to assess the model's ability to learn from datasets with small to medium sample sizes per class, we performed hyperparameter tuning using grid search and k-fold cross-validation for each scenario (20 spc, 500 spc, and ``full dataset''). This means that for each fold, a total of $160$ combinations were tested for the 20 spc, 500 spc, and ``full dataset''. In contrast, due to the considerable size of the ELAsTiCC dataset, we opted to execute hyperparameter tuning with just one validation fold. This decision stemmed from the high computational expense associated with evaluating multiple folds and the anticipated low variance across folds in large-scale datasets. As before, a total of $160$ combinations were tested, but only in one fold. The lower and upper bounds for the hyperparameters marker size and line width were determined based on the degree of visual overlap between the observations and lines in the images. Figure~\ref{fig:input_model} (a) and (b) present images generated using the best hyperparameters identified for the ``Grid'' and ``Overlay'' approaches, respectively.

To mitigate class imbalance, we employed a weighted sampling strategy\footnote{\protect\href{https://pytorch.org/docs/stable/data.html\#torch.utils.data.WeightedRandomSampler}{PyTorch's WeightedRandomSampler.}} that adjusted the selection probability of each class according to its relative frequency, promoting a more balanced representation in training and improving model generalization. Additionally, we optimized the model using Adam and applied early stopping with a patience of 10 epochs for the MACHO datasets and 5 epochs for ELAsTiCC, based on the validation F1-score.

\begin{table}[t]
\centering
\caption{Summary of the hyperparameters used for model optimization.}
\label{table:hyperparameters} 
\renewcommand{\arraystretch}{1.35}
\begin{tabular}{ll}
\hline\hline
Hyperparameter & Values                 \\ \hline
Marker size     & \{1, 2, 3, 4, 5\}      \\
Line width      & \{0.5, 1.0, 1.5, 2.0\} \\
Flux errors    & \{True, False\}            \\
Input format   & \{Grid, Overlay\}      \\
Learning rate  & \{$5\cdot10^{-5}$, $5\cdot10^{-6}$\}         \\ \hline
\end{tabular}
\end{table}

\section{Results}\label{sec:results}
We evaluated the classification performance of the SwinV2 model on two previously mentioned datasets: MACHO and ELAsTiCC. These datasets exhibit distinct characteristics, particularly in terms of the number of observations per light curve, the number of bands, and their cadences and taxonomies, which influenced both hyperparameter choices and classification performance. We initiate our analysis with the MACHO dataset, exploring both single-band and two-band experiments, before expanding our assessment to the ELAsTiCC dataset, which includes six bands. We trained a separate model for each fold and report the macro F1-score as the mean and standard deviation computed across the models evaluated on the test set, referred to in the text as simply the F1-score.

\begin{table}[b]
    \caption{F1-score (\%) on the MACHO test set for models trained with 20 and 500 samples per class (spc), and the ``full dataset'' comprising $20\,294$ labeled light curves used for training.}
    \label{table:f1_scores_swinv2_astromer} 
    \centering
    \renewcommand{\arraystretch}{1.35}
    \begin{tabular}{lccc}
        \hline\hline
        Model & 20 spc & 500 spc & Full dataset \\ 
        \hline
        Astromer - F & 56.0 $\pm$ 2.5  & 71.0 $\pm$ 1.8  & - \\
        Astromer - T & 44.0 $\pm$ 5.0  & 73.0 $\pm$ 3.0  & - \\
        SwinV2 - T   & \textbf{60.1 $\pm$ 4.6}  & \textbf{77.4 $\pm$ 1.3}  & \textbf{79.8 $\pm$ 2.2} \\
        \hline
    \end{tabular}
    \tablefoot{F denotes a frozen transformer, where only the classifier is updated. T represents a trainable transformer, where both the transformer and classifier are updated during training.}
\end{table}

\subsection{Model performance in single- and multi-band light curves}
To evaluate the classification performance of our approach using a single band, we compared SwinV2 with Astromer. Our initial experiments focused on the R band of the MACHO dataset to ensure direct comparability with the results reported in the referenced study. We conducted hyperparameter tuning to optimize the model's performance and to assess its ability to learn from datasets with small to medium sample sizes per class. The best combination of hyperparameters, using one band, for the 20 and 500 spc used to train the model, as well as the ``full dataset'', can be found at the top of Table~\ref{table:best_hp} in Appendix \ref{sec:best_hp}. These combinations were then used to generate the SwinV2 results presented in Table~\ref{table:f1_scores_swinv2_astromer}.

\begin{figure}[b]
  \resizebox{\hsize}{!}{\includegraphics{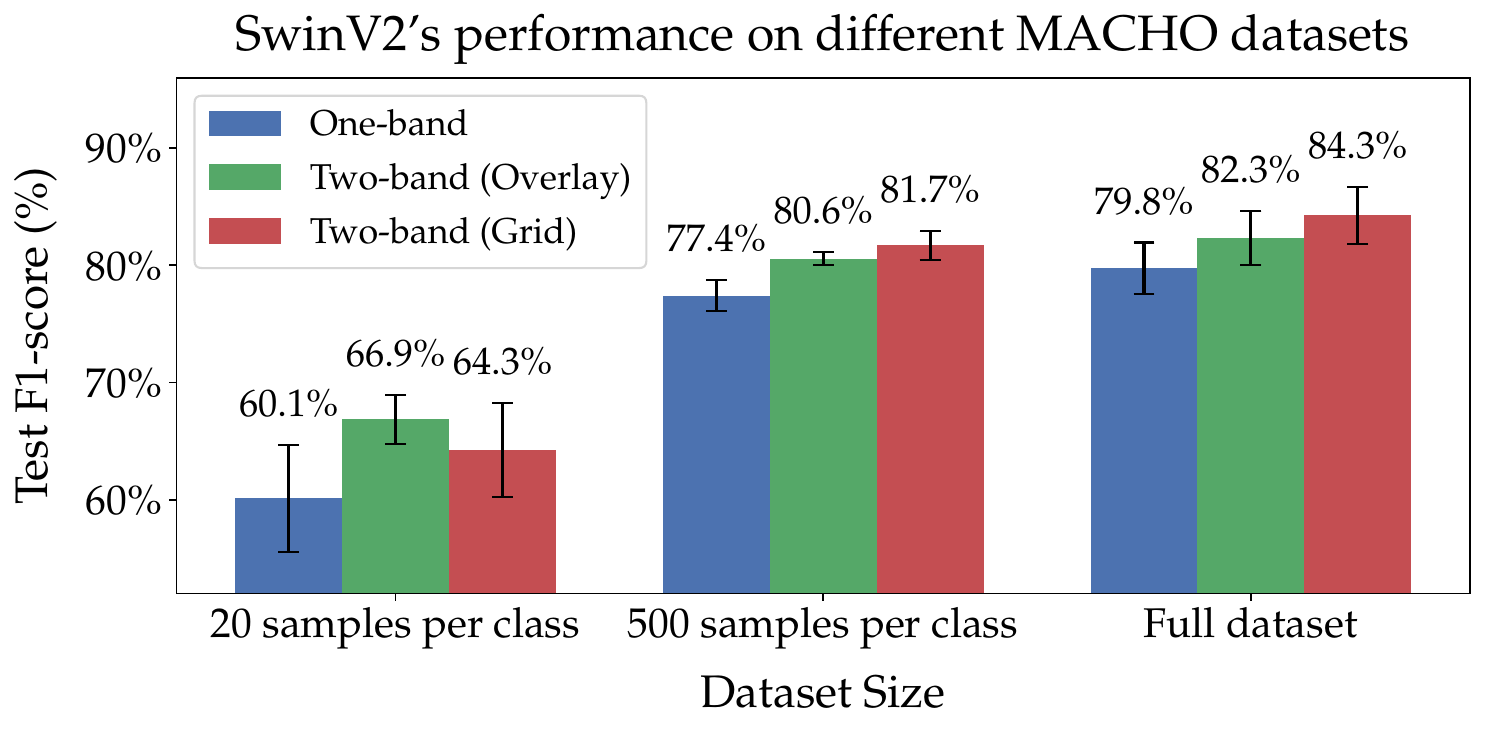}}
  \caption{SwinV2 classification performance on the MACHO test set using single- and multi-band data trained on varying sample sizes. The figure presents the best results from the hyperparameter search for both the ``Overlay'' and ``Grid'' approaches.}
  \label{fig:f1_scores_swinv2_bands}
\end{figure}

Table~\ref{table:f1_scores_swinv2_astromer} presents the single-band classification results on the MACHO dataset in the R band for SwinV2 and Astromer. The table reports the F1-scores along with their standard deviations for SwinV2 (in trainable configurations) and Astromer (in both frozen and trainable configurations) across different dataset sizes. In this context, a trainable model refers to one in which both the transformer and classifier layers are updated during training, whereas a frozen model retains the transformer's pre-trained weights and only updates the classifier. Since results for the ``full dataset'' were not available for Astromer, we compared it only in the limited-data scenarios (20 and 500 spc). However, the ``full dataset'' results provide a basis for evaluating our multi-band method. As shown in Table~\ref{table:f1_scores_swinv2_astromer}, SwinV2 consistently outperformed Astromer across all configurations and dataset sizes. For 20 spc, SwinV2 achieved an F1-score of $60.1\%$, which is $4.1\%$ points higher than Astromer's best result of $56.0\%$. For 500 spc, SwinV2 attained $77.4\%$, surpassing Astromer by $4.4\%$ points. These results demonstrate the robustness and adaptability of the VT model in achieving high classification performance on light curves, particularly in low-data scenarios. When tested on the unbalanced dataset (i.e., the ``full dataset'', as illustrated in Fig.~\ref{fig:distribution_macho}), SwinV2 achieved an F1-score of $79.8\%$, surpassing its performance on the balanced dataset with 500 spc ($77.4\%$). This result highlights the VT model's ability to maintain strong performance even in the presence of data imbalance.

\begin{figure}[t]
    \centering
    \includegraphics[width=\hsize]{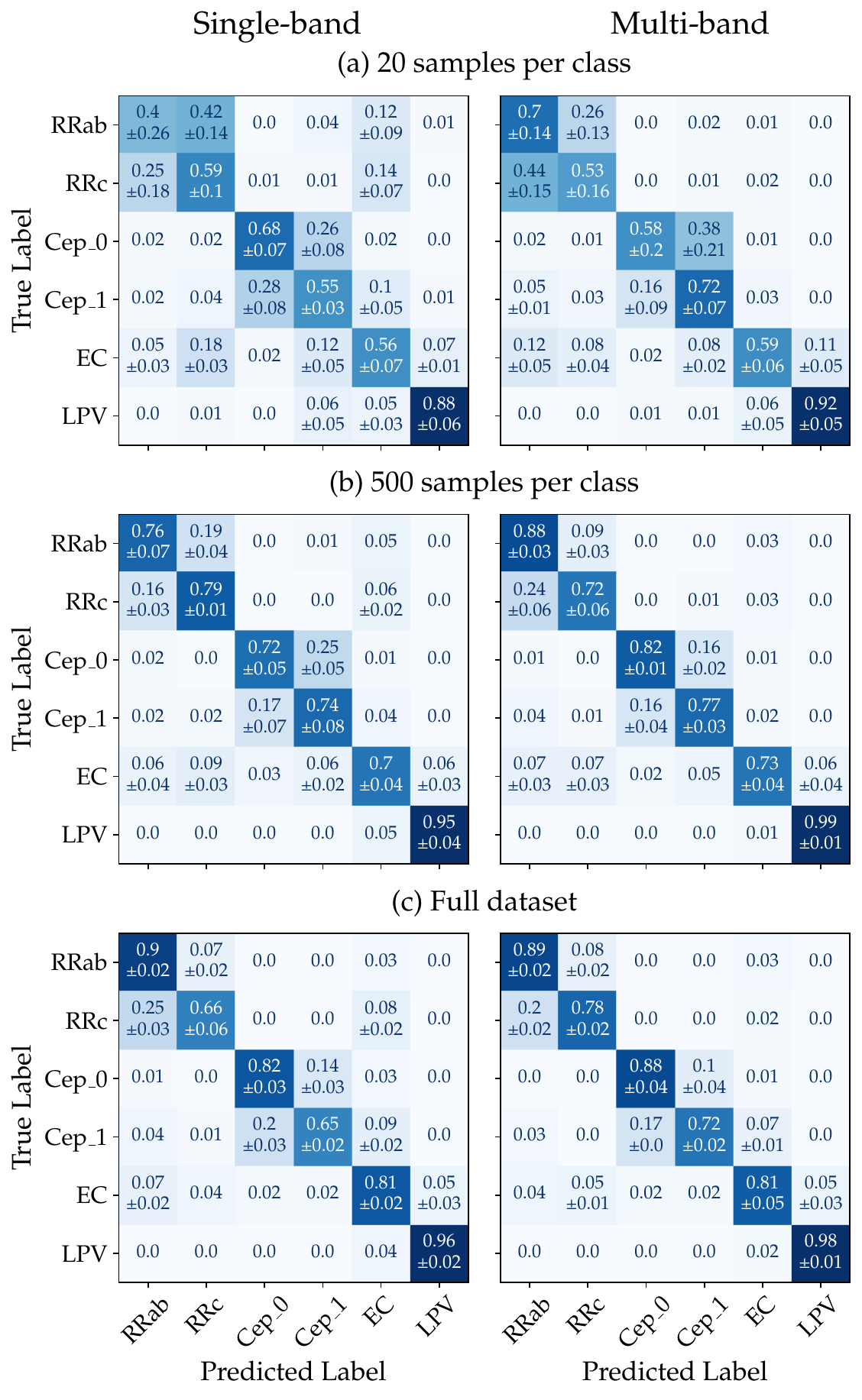}
    \caption{Confusion matrices for the best classification results obtained using SwinV2 with single-band and two-band inputs on different subsets of the MACHO dataset. Standard deviations are displayed for values greater than or equal to $0.05$.}
    \label{fig:cm_macho}
\end{figure}

To demonstrate the impact of the multi-band approach, we extended our analysis to incorporate both the R and B bands from the MACHO dataset, employing two distinct plotting strategies: ``Grid'' and ``Overlay''. As before, we performed hyperparameter tuning on each scenario. The best combination of hyperparameters using two bands can be seen in the middle of Table~\ref{table:best_hp} in Appendix \ref{sec:best_hp}. The results in Fig.~\ref{fig:f1_scores_swinv2_bands} show that incorporating multi-band inputs improves F1-scores compared to single-band experiments. Using two bands led to performance improvements of $6.8\%$, $4.3\%$, and $4.5\%$ for the 20, 500, and ``full dataset'' settings, respectively, relative to the best multi-band approach in each case. These improvements highlight the model's ability to effectively leverage the additional information provided by multiple bands, reinforcing the value of multi-band configurations for capturing the complex behavior of light curves through image-based representations. For 20 spc, ``Overlay'' outperformed ``Grid'' (F1-score: $66.9\%$ vs. $64.3\%$). With 500 spc, ``Grid'' achieved an F1-score of $81.7\%$, marginally surpassing ``Overlay'' ($80.6\%$). When using the ``full dataset'', ``Grid'' attained a higher F1-score of $84.3\%$, slightly outperforming ``Overlay'' ($82.3\%$). These results indicate that ``Grid'' achieved better performance in two configurations. However, the differences were minimal.

The confusion matrices for the best-performing SwinV2 models across different dataset sizes, for both single- and multi-band settings, are shown on the left and right of Fig.~\ref{fig:cm_macho}, respectively. These matrices illustrate the recall per class, providing insight into how well the model correctly identifies each category. For the subset with 20 spc, using two bands instead of one improved recall in 4 out of 6 classes (RRab, Cep\_1, EC, and LPV) but decreased it for RRc and Cep\_0. For the subset with 500 spc, recall increased in 5 out of 6 classes (RRab, Cep\_0, Cep\_1, EC, and LPV) but decreased by $7\%$ for RRc. For the ``full dataset'', a similar pattern was observed. Recall improved in 4 out of 6 classes (RRc, Cep\_0, Cep\_1, and LPV), remained unchanged for EC, and decreased by $1\%$ for RRab. The consistent improvement observed in Cep\_1, EC, and LPV suggests that these classes benefit from the additional spectral information. In contrast, RRab, RRc, and Cep\_0 exhibit mixed results, indicating that while multi-band inputs are beneficial in most cases, they do not always provide a clear advantage, especially for certain classes that remain challenging to distinguish due to their similar flux variations in both bands. This is particularly evident in RRc and RRab, as well as Cep\_0 and Cep\_1.

\begin{figure}[t]
  \resizebox{\hsize}{!}{\includegraphics{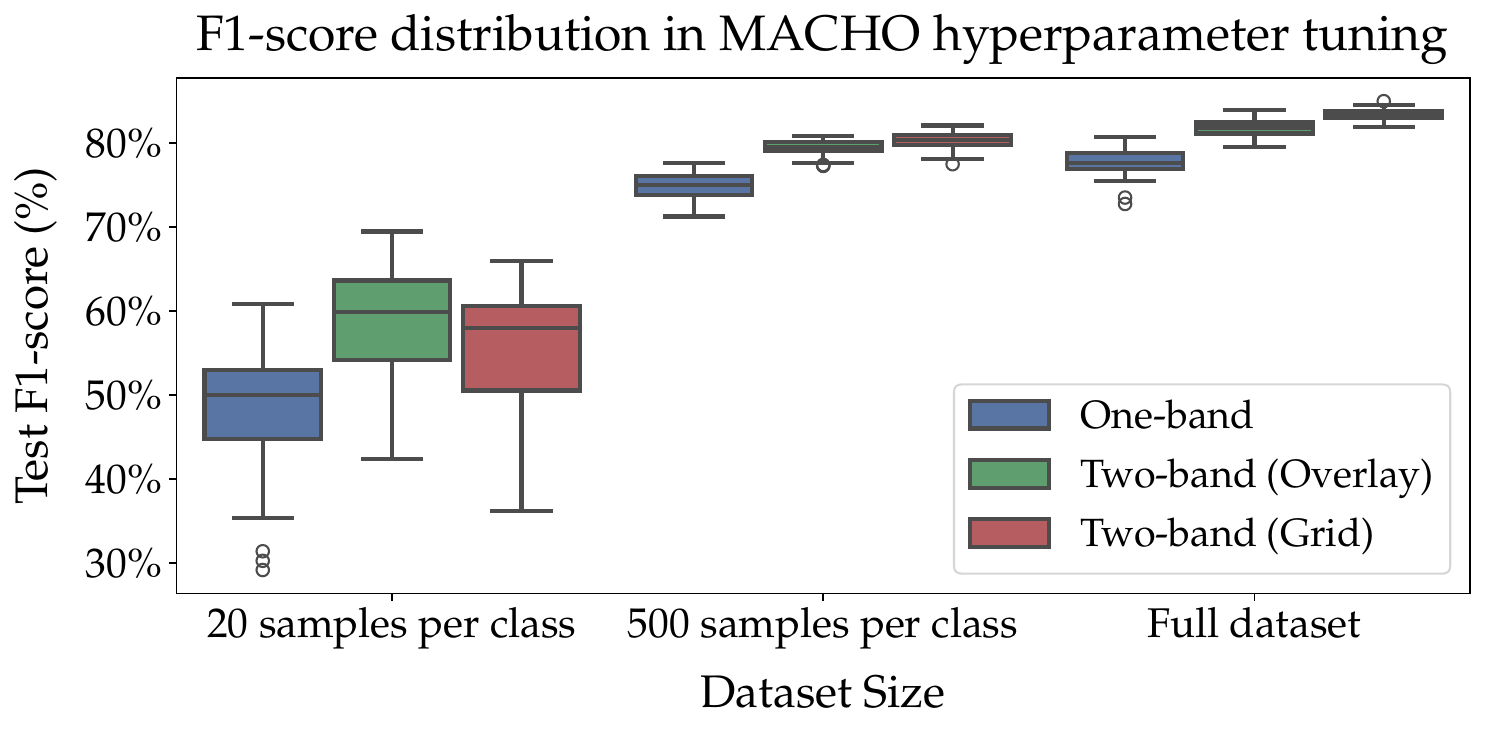}}
  \caption{Distribution of F1-scores obtained during hyperparameter tuning for the MACHO dataset comparing single-band and two-band (``Overlay'' and ``Grid'') approaches across different dataset sizes (20 spc, 500 spc, and the ``full dataset'').}
  \label{fig:f1_hp_macho}
\end{figure}

Figure~\ref{fig:f1_hp_macho} shows the distribution of F1-scores obtained during the hyperparameter tuning process for the MACHO dataset, considering both single-band and two-band approaches for each scenario (20 spc, 500 spc, and the ``full dataset''). In general, the median F1-score values across all combinations remain consistent with those of the best combinations presented in Fig.~\ref{fig:f1_scores_swinv2_bands}, where the ``Overlay'' approach performed better for 20 spc, while the ``Grid'' approach was superior for 500 spc and the ``full dataset''. In particular, the F1-score for 500 spc and the ``full dataset'' exhibited little variation, regardless of the hyperparameter combinations used. The exception is the 20 spc scenario, where a poor choice of hyperparameters can lead to low performance due to the limited amount of data, which restricts the model's ability to extract sufficient discriminative information from the light curves.

Evaluating model performance under realistic observational constraints is important to understand how models behave beyond controlled experimental settings. In our earlier experiments (see Fig.~\ref{fig:f1_scores_swinv2_bands}), we used randomly sampled subsets of the MACHO dataset with 20 and 500 samples per class to compare SwinV2 with Astromer. These results showed that SwinV2 reached higher F1-scores across different training sizes. However, those experiments did not reflect the sampling biases often found in real spectroscopic follow-up observations, where brighter sources tend to be prioritized.

\begin{table}[b]
    \caption{F1-score (\%) on the MACHO test set using the best SwinV2 model, evaluated under brightness-biased sampling in the 20 spc-b and 500 spc-b settings. The test set remains unchanged.}
    \label{table:f1_scores_swinv2_brightbiased} 
    \centering
    \renewcommand{\arraystretch}{1.35}
    \begin{tabular}{lcc}
        \hline\hline
        Approach & 20 spc-b & 500 spc-b \\ 
        \hline
        One-band & 48.9 $\pm$ 1.4 & 69.5 $\pm$ 1.8 \\
        Two-band (Overlay) & \textbf{55.8 $\pm$ 4.9} & 72.9 $\pm$ 1.5 \\
        Two-band (Grid)    & 51.6 $\pm$ 7.0 & \textbf{74.7 $\pm$ 1.2} \\
        \hline
    \end{tabular}
\end{table}

To explore this aspect, we conducted a complementary experiment using a brightness-biased sampling strategy. Specifically, we modified only the training set by selecting the brightest objects of each class from the ``full dataset'', based on their median magnitudes, while maintaining the 20 and 500 samples per class configurations. We refer to these settings as 20 spc-b and 500 spc-b, where the suffix -b denotes brightness-biased sampling. The test set remained unchanged from previous experiments. This introduces a distribution shift between training and testing, as the training set includes only the brightest sources, while the test set covers the full magnitude range. The results, presented in Table~\ref{table:f1_scores_swinv2_brightbiased}, show lower F1-scores compared to those obtained with randomly sampled training data. For 20 spc-b, the F1-score dropped from 60.1\% to 48.9\% in the one-band configuration, from $66.9\%$ to $55.8\%$ in the ``Overlay'' approach, and from $64.3\%$ to $51.6\%$ in the ``Grid'' approach. For 500 spc-b, the F1-score decreased from $77.4\%$ to $69.5\%$ in one-band, from $80.6\%$ to $72.9\%$ in ``Overlay'', and from $81.7\%$ to $74.7\%$ in ``Grid''. Despite these differences, the model still achieves relatively high scores, particularly with two-band inputs. Among the visual strategies, ``Overlay'' performs better under smaller training sizes, while ``Grid'' yields the highest performance with 500 spc-b. These findings indicate that the model remains effective even under sampling conditions that reflect realistic observational biases.

\subsection{Model performance in ELAsTiCC } \label{sec:ELAsTiCCresults}

\begin{table}[t]
\centering
\caption{F1-score (\%) on the ELAsTiCC test set for ATAT using only light curves and for SwinV2 with the ``Overlay'' and ``Grid'' approaches.}
\label{table:f1_scores} 
\renewcommand{\arraystretch}{1.35}
\begin{tabular}{lc}
\hline\hline
Model              & F1-score \\ \hline
ATAT (LC)     & 64.2 $\pm$ 0.69           \\
SwinV2 (Overlay)               & \textbf{65.5 $\pm$ 0.28}           \\ 
SwinV2 (Grid)                 & 64.6 $\pm$ 0.22           \\ \hline
\end{tabular}
\end{table}

\begin{figure*}[b]
    \centering
    \includegraphics[width=\hsize]{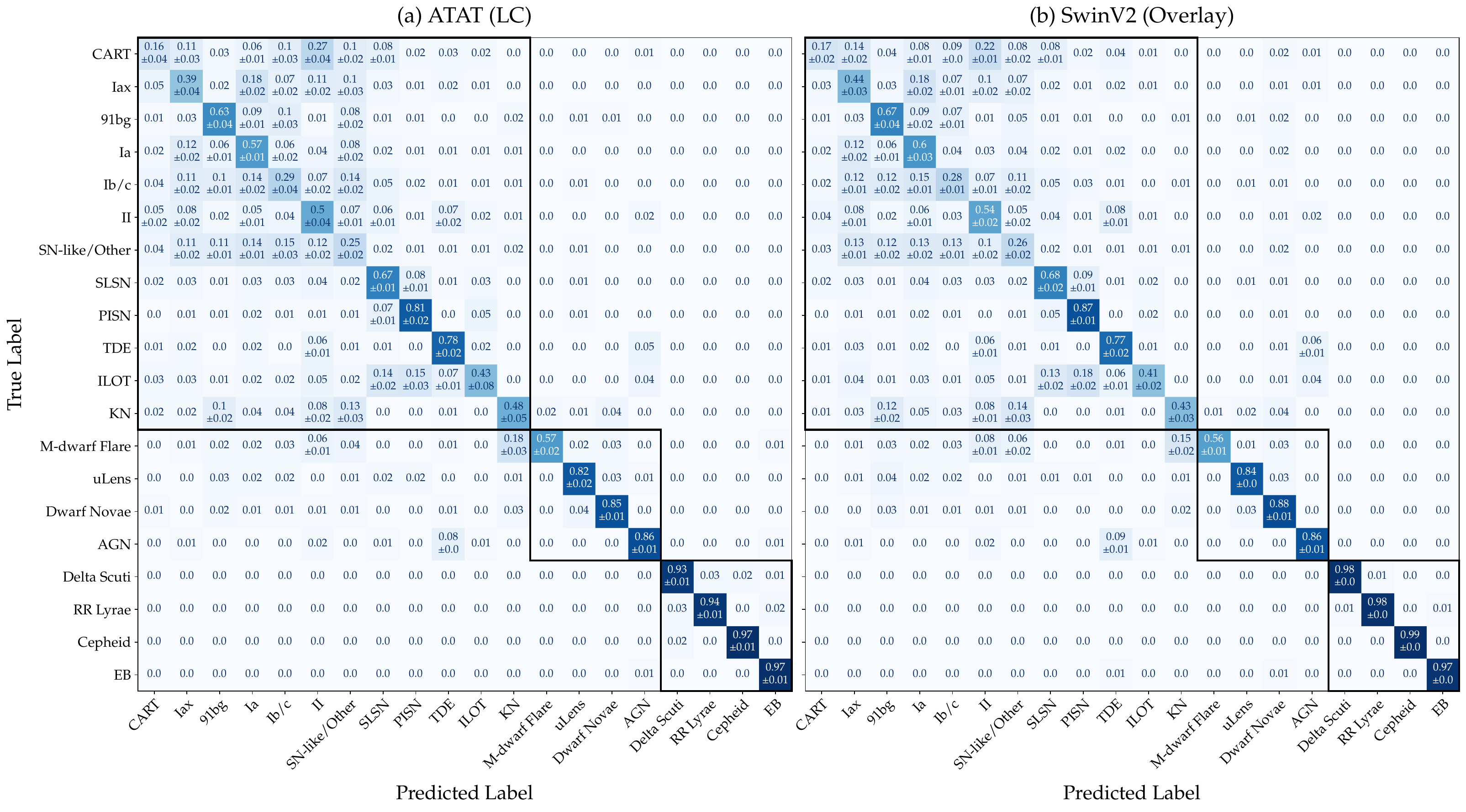}
    \caption{Confusion matrices for the ATAT model trained exclusively on light curves and the best-performing SwinV2 model, both of which use the ELAsTiCC dataset. The black lines delineate the three main groups of objects used in the ATAT paper: the transient, stochastic, and periodic groups, from left to right, respectively. Standard deviations are displayed for values greater than or equal to $0.05$.}
    \label{fig:cm_atat_swinv2}
\end{figure*}

To assess the model’s performance on a large-scale dataset, we used ELAsTiCC, which includes six bands. In line with prior datasets, we fine-tuned hyperparameters to enhance the model's effectiveness. The best hyperparameter combination for ``Grid'' and ``Overlay'' can be found at the bottom of Table~\ref{table:best_hp} in Appendix \ref{sec:best_hp}.

Table~\ref{table:f1_scores} presents the F1-scores along with their standard deviations for SwinV2’s best results, compared to the  Astronomical Transformer for time series And Tabular data \citep[ATAT;][]{cabrera2024atat}, a specialized time domain model designed for multi-band light curves. Since ATAT provides results that combine light curves with metadata and additional features, we focused on its light curve-based results to ensure a fair comparison and demonstrate the effectiveness of our approach on equal footing. We refer to this as ATAT (LC).  The results show that SwinV2 slightly outperformed ATAT (LC) in both configurations, with the ``Overlay'' approach achieving the highest F1-score of $65.5\%$, followed by the ``Grid'' approach with $64.6\%$, while ATAT (LC) obtained $64.2\%$. Although the $1.3\%$ improvement of SwinV2 (Overlay) over ATAT (LC) is moderate, it underscores the model’s ability to effectively distinguish between 20 transient and variable star classes in a more complex classification task.

\begin{figure}[t]
  \resizebox{\hsize}{!}{\includegraphics{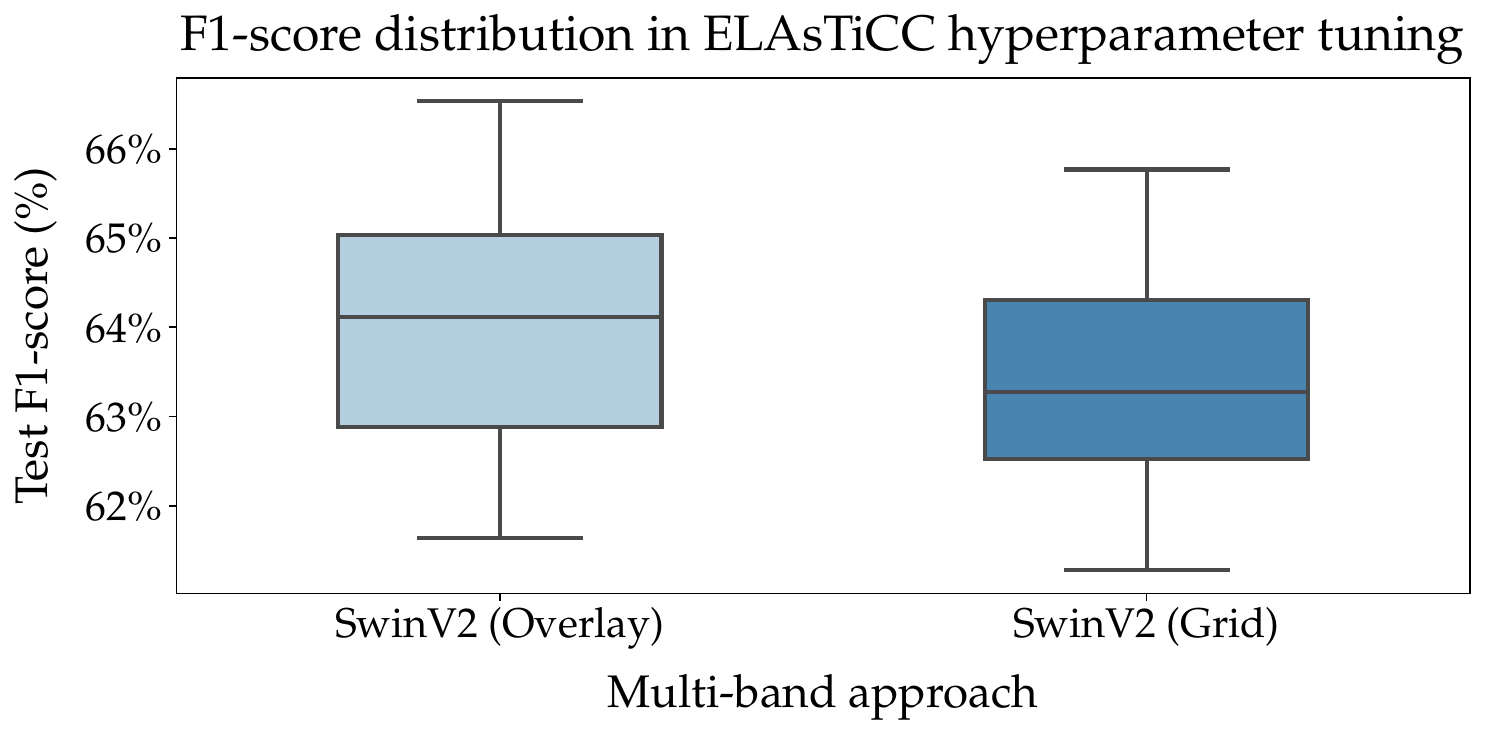}}
  \caption{Distribution of F1-scores obtained during hyperparameter tuning for the ELAsTiCC dataset comparing the ``Overlay'' and ``Grid'' multi-band approaches.}
  \label{fig:f1_hp_elasticc}
\end{figure}

Figure~\ref{fig:cm_atat_swinv2} presents the confusion matrices for the ATAT (LC) model and the SwinV2 (Overlay) model, which achieved the best performance on the ELAsTiCC dataset. Since the confusion matrix for ATAT (LC) was not directly available, we generated it using the publicly available prediction files from the referenced study\footnote{\href{https://github.com/alercebroker/ATAT?tab=readme-ov-file\#important-files}{Files with the results of the ATAT paper.}}. In general terms, a similar pattern can be observed in Fig.~\ref{fig:cm_atat_swinv2} (a) and (b), where both models tend to misclassify the same classes and correctly identify light curves in similar proportions. Both primarily struggle to differentiate between various supernova types and achieve high recall in the stochastic group (ATAT: $\geq 82\%$, SwinV2: $\geq 84\%$, both except for M-dwarf Flare) and the periodic group (ATAT: $\geq 93\%$, SwinV2: $\geq 97\%$). SwinV2 (Overlay) slightly outperformed ATAT (LC) in 13 out of 20 classes, obtained the same results in 2 classes, and showed a slight decrease in 5 classes. Specifically, SwinV2 (Overlay) demonstrated higher recall in 8 out of 12 supernova subclasses, namely: CART, Iax, 91bg, Ia, II, SN-like/Other, SLSN, and PISN, with improvements of more than $4\%$ in the Iax, 91bg, II, and PISN types. Additionally, SwinV2 (Overlay) performed equally well or better in 3 out of 4 stochastic subclasses and in all 4 periodic subclasses, where its recall improved by more than $4\%$ for the Delta Scuti and RR Lyrae classes.

Figure~\ref{fig:f1_hp_elasticc} illustrates the distribution of F1-scores obtained during the hyperparameter tuning process for the ELAsTiCC dataset, comparing the ``Overlay'' and ``Grid'' approaches. As before, the median F1-score values across all tested hyperparameter combinations remain consistent with those of the best configurations reported in Table~\ref{table:f1_scores}, where SwinV2 (Overlay) slightly outperforms SwinV2 (Grid). The F1-score variations across different hyperparameter settings fall within a narrow range of $61\%$ to $67\%$,  indicating that classification performance remains relatively stable regardless of the chosen hyperparameters.

In comparison to MACHO, which consists primarily of periodic stars, ELAsTiCC presents a more complex taxonomy, making the classification task more challenging. In both datasets, SwinV2 demonstrated strong performance in distinguishing periodic stars, and in ELAsTiCC, it also proved effective at identifying stochastic stars. As mentioned earlier, the main difficulty arises in the classification of supernovae, where both SwinV2 and ATAT struggled to differentiate between subclasses. In \citet{cabrera2024atat}, ATAT was also evaluated in a configuration that incorporated both light curves and metadata (tabular information provided in the alert stream), referred to as ATAT (LC + MD). The inclusion of metadata significantly improved its performance, increasing the F1-score from $64.2\%$ (LC) to $82.6\%$ (LC + MD), which emphasizes the importance of metadata or additional features in improving supernova classification. Exploring the integration of tabular data in SwinV2 is left for future work, as it could yield similar improvements to those observed in ATAT.

\section{Conclusions}\label{sec:conclusions}

We have introduced a simple and effective method for classifying photometric light curves using a pre-trained VT model without the need for complex feature engineering or multi-band preprocessing to handle multi-band observations. We demonstrated the capability and effectiveness of this method using one, two, and six bands of light curves as well as scenarios involving imbalanced datasets, limited data, and the classification of both transient and variable stars.

Our results show that leveraging the knowledge acquired from pre-training SwinV2 on a large dataset and fine-tuning it for a specific task can lead to better performance than models specifically designed for light curve data. When using a single band in the MACHO dataset, we demonstrated that, compared to Astromer, our method improves the F1-score by $4.1\%$ in a dataset with 20 spc used to train the model and by $4.4\%$ in a dataset with 500 spc. Additionally, we showed that it is possible to maintain a high performance ($79.8\%$) of F1-score on the unbalanced ``full dataset''. When incorporating a second band in the same dataset, we observed further improvements compared to SwinV2 with one band: an additional $6.8\%$ in the F1-score for 20 spc, $4.3\%$ for 500 spc, and $4.5\%$ in the ``full dataset''. Finally, when evaluating our method on the ELAsTiCC dataset, we demonstrated that it achieves an F1-score of $65.5\%$, which is $1.2\%$ higher than the score obtained in ATAT using only light curves.

This study highlights the discriminative power of pre-trained VT models and demonstrates their ability to adapt to datasets that differ entirely from those on which they were originally trained. By analyzing images of light curves, the model successfully distinguished between astronomical objects across datasets with distinct observational characteristics, such as the MACHO survey and ELAsTiCC. A key factor in this success is the model’s ability to capture the temporal behavior of light curves, which is primarily achieved through the positional embedding. This embedding provides a structured representation of the light curve’s spatial arrangement within the image, allowing the model to implicitly encode ordering information. As a result, it delivers competitive performance against models specifically designed to process sequential light curves and account for the distances between observations. This capability not only underscores the model’s versatility but also suggests a promising direction for developing a more generalizable approach. In particular, it could serve as a foundation for a model capable of adapting to any type of survey with minimal fine-tuning using only a few light curves. However, further investigation is necessary to fully validate this potential.

A direct extension of our approach would be to incorporate metadata alongside light curves to provide additional context and enhance classification performance, which should align with the improvements observed in ATAT when using both light curves and metadata (LC + MD). Additionally, pre-training the model on light curves could be explored to assess its ability to extract general information from astronomical time series using a self-supervised task. Future work could explore integrating descriptive information, such as textual captions for each light curve. If such data were available, we could extend our method to vision-language transformers (VLTs), enabling tasks such as image retrieval (IR) with CLIP \citep{radford2021learning}, visual question answering (VQA) with BLIP-2 \citep{li2023blip}, and even light curve image deblurring using Stable Diffusion \citep{rombach2022high}. Given the potential computational cost of these models, knowledge distillation could be employed to improve inference efficiency while maintaining accuracy. These advancements would further refine the classification of photometric light curves and open new avenues for multi-modal astronomical analysis in the time domain.

\section*{Code and data availability}
The code used in this study, along with instructions on how to access the data, is available at \href{https://github.com/dnlmoreno/VT_Model_for_LightCurves_Classification}{GitHub repository}. Detailed information on how to obtain and use the dataset is provided in the README file.

\begin{acknowledgements}
      The computations in this paper were run on the FASRC Cannon cluster supported by the FAS Division of Science Research Computing Group at Harvard University.

      The authors acknowledge support from the National Agency for Research and Development (ANID) grants: Millennium Science Initiative ICN12\_009 and AIM23-0001 (GCV), NCN2021\_080 (GCV, CDO), and FONDECYT Regular 1231877 (DMC, GCV, MCL, CDO).
\end{acknowledgements}

\bibliographystyle{aa}
\bibliography{reference}

\begin{appendix}

\onecolumn 
\section{Visual representations of MACHO and ELAsTiCC light curves}
\label{appendix:visual_samples}

\begin{figure*}[h!]
  \centering
  \includegraphics[width=\linewidth]{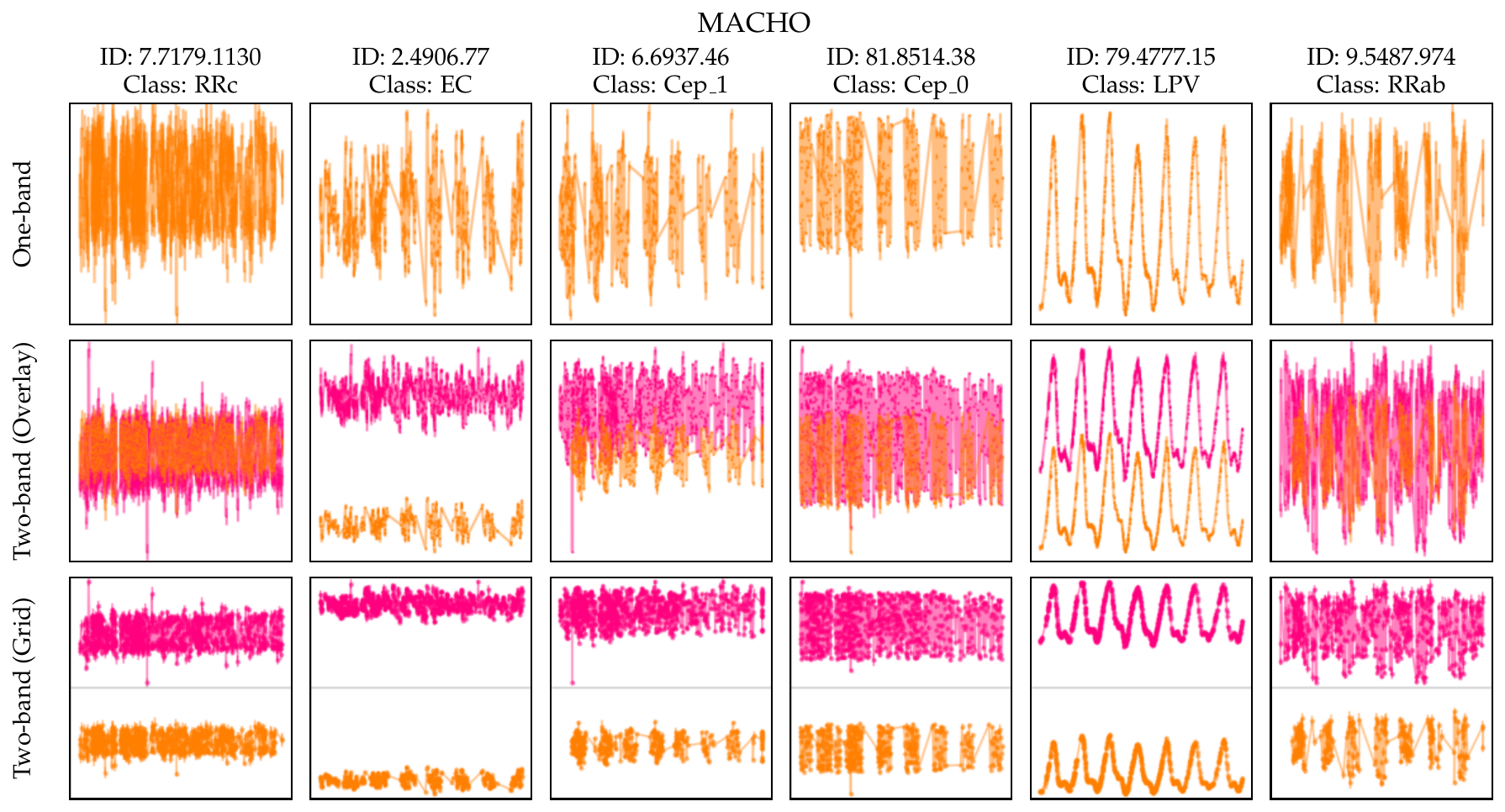}
  \caption{Examples from the MACHO dataset illustrating six objects from different classes. Each object is shown in three formats: the original single-band light curve and two-band representations generated using the ``Overlay'' and ``Grid'' approaches.}
  \label{fig:random_samples_macho}
\end{figure*}

\begin{figure*}[h!]
  \centering
  \includegraphics[width=\linewidth]{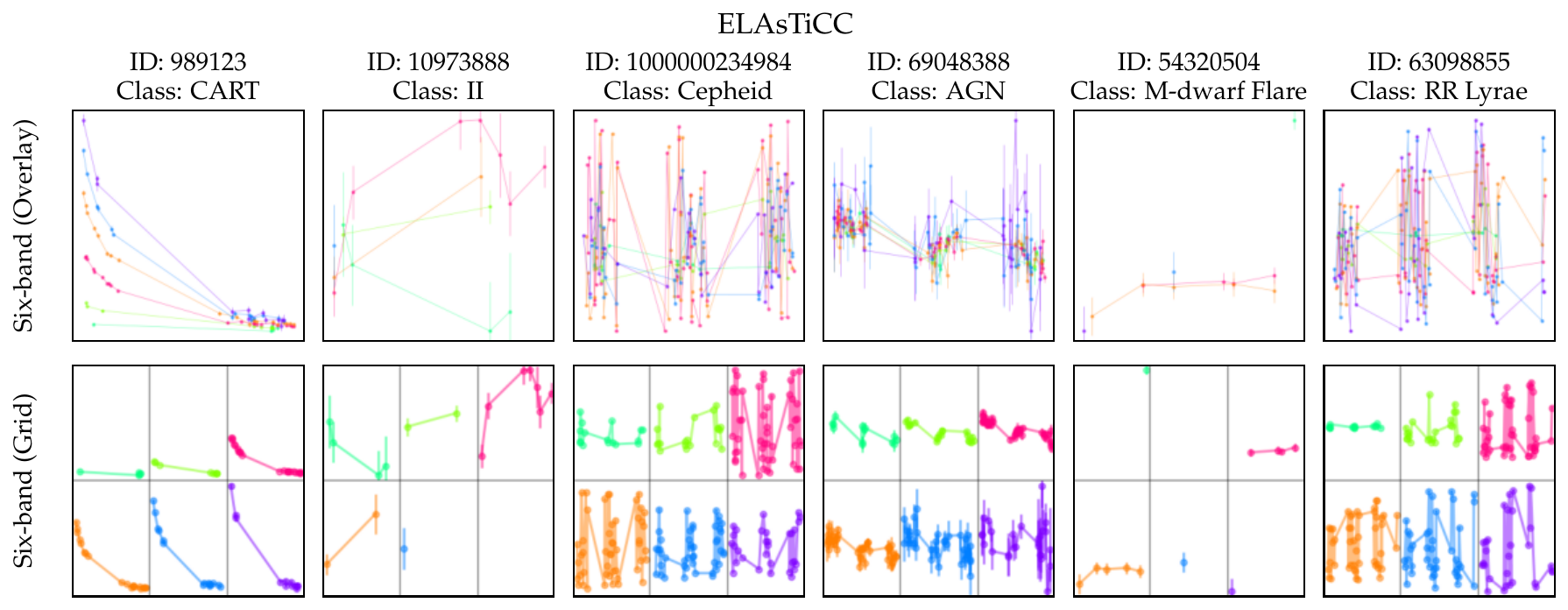}
  \caption{Examples from the ELAsTiCC dataset showing six-band light curves for six objects from different classes. Each object is represented using both the ``Overlay'' and ``Grid'' approaches.}
  \label{fig:random_samples_elasticc}
\end{figure*}

Figures~\ref{fig:random_samples_macho} and~\ref{fig:random_samples_elasticc} present randomly selected examples from the MACHO and ELAsTiCC datasets to illustrate the visual characteristics of the light curves used in our experiments. Each figure shows six objects from different classes, represented using the approaches evaluated in our pipeline and visualized with the hyperparameters that yielded the best classification performance. For MACHO, we include three visual representations per object: the single-band light curve (One-band), and the two-band inputs constructed using the ``Overlay'' and ``Grid'' approaches. For ELAsTiCC, which contains six-band light curves, we show the corresponding ``Overlay'' and ``Grid'' representations. Within each dataset, the same objects and class labels are used across approaches to enable direct visual comparison of how different methods encode temporal and spectral information.

\FloatBarrier
\twocolumn

\onecolumn
\section{Hyperparameters}
\label{sec:best_hp}

Table~\ref{table:best_hp} presents the best hyperparameters obtained for marker size, line width, flux errors, input format, and learning rate for the MACHO dataset (single-band and two-band) in the scenarios of 20 spc, 500 spc, and the ``full dataset'', as well as for the ELASTiCC dataset (six-band). Hyperparameter tuning was conducted using a grid search over the values listed in Table~\ref{table:hyperparameters}.

The results reveal a few key patterns: the optimal marker size and line width depend on both the number of bands and the approach used to represent multi-band observations. When using the ``Grid'' approach, a larger marker size is necessary because each light curve band is represented on a lower-resolution grid, requiring larger observation markers for better visibility. This effect is particularly evident in ELAsTiCC, where the grid is divided into six sections. The same pattern is observed in the ``full dataset'' of MACHO (two-band) but not in the 20 spc and 500 spc scenarios, where the limited number of observations increases the likelihood of weak data representation. Conversely, the ``Overlay'' approach generally exhibits less variation in the marker size and line width hyperparameters across the different ``full datasets'', including MACHO single-band, MACHO two-band, and ELAsTiCC, remaining within a marker size range of 1.0 to 2.0 and a line width range of 0.5 to 2.0. This consistency occurs because the resolution at which bands are represented remains fixed at $256 \times 256$. In the same context, flux errors appear to be important in almost all cases, except in the 500 spc scenario for MACHO (two-band). Additionally, higher learning rates performed better for smaller datasets, while lower learning rates were more effective for larger datasets. This trend likely occurs because higher learning rates help escape local optima when data is scarce, whereas lower learning rates enable more stable updates when training on larger datasets.

\begin{table*}[h!]
\centering
\caption{Summary of the best hyperparameters found for the SwinV2 model across single- and multi-band cases and varying dataset sizes.}
\label{table:best_hp} 
\renewcommand{\arraystretch}{1.35}
\begin{tabular}{lccccc}
\hline\hline
\textbf{MACHO (single-band)}  & Marker size & Line width & Flux errors & Input format    & Learning rate \\ \hline
20 spc               & 5.0        & 1.5       & True  & Overlay & $5\cdot10^{-5}$ \\
500 spc              & 1.0        & 1.0       & True  & Overlay & $5\cdot10^{-6}$ \\
Full dataset         & 1.0        & 2.0       & True  & Overlay & $5\cdot10^{-6}$ \\ \hline
\textbf{MACHO (two-band)}    & Marker size & Line width & Flux errors & Input format & Learning rate \\ \hline
20 spc               & 1.0        & 2.0       & True  & Grid    & $5\cdot10^{-5}$ \\
20 spc               & 2.0        & 0.5       & True  & Overlay & $5\cdot10^{-5}$ \\
500 spc              & 2.0        & 1.0       & False & Grid    & $5\cdot10^{-5}$ \\
500 spc              & 3.0        & 1.0       & True  & Overlay & $5\cdot10^{-5}$ \\
Full dataset         & 3.0        & 1.5       & True  & Grid    & $5\cdot10^{-6}$ \\
Full dataset         & 1.0        & 2.0       & True  & Overlay & $5\cdot10^{-6}$ \\ \hline
\textbf{ELAsTiCC (six-band)} & Marker size & Line width & Flux errors & Input format & Learning rate \\ \hline
Full dataset         & 5.0        & 2.0       & True  & Grid    & $5\cdot10^{-6}$ \\
Full dataset         & 2.0        & 0.5       & True  & Overlay & $5\cdot10^{-6}$ \\ \hline
\end{tabular}
\end{table*}

\end{appendix}

\end{document}